\documentclass{iau} 
\usepackage{natbib}
\usepackage{graphicx}

\title[JD 11.~~Evolution of Granular Asteroids] 
{Asteroid Evolution: \\Role of Geotechnical Properties}

\author[Paul S{\'a}nchez]   
{Paul S{\'a}nchez$^1$}
 
\affiliation{$^1$Colorado Center for Astrodynamics Research, University of Colorado Boulder, Boulder, CO 80309-0431\\ email: {\tt diego.sanchez-lana@colorado.edu}}

\pubyear{2015}
\volume{318}  
\jname{Asteroids: New Observations, New Models}
\editors{Steven Chesley, Alessandro Morbidelli, \\ Robert Jedicke \& Davide Farnocchia, eds.}
\begin{document}

\maketitle

\begin{abstract}
This paper presents a brief review and latest results of the work that has been carried out by the Planetary Science community in order to understand that role of the geotechnical properties of granular asteroids (commonly known as ``rubble-pile'' asteroids) in their formation, evolution and possible disruption.  As such, we will touch in aspects of the theoretical and numerical tools that have been used with this objective and how the obtained results compare to the observed asteroids.  
\end{abstract}

\firstsection 
\section{Introduction}

Over the last decade of research in Planetary Science, the scientific community has made many advances in their understanding of the evolution of asteroids in the Solar System. One particular area of fruitful study started with the bold idea that these small planetary bodies could be gravitational aggregates and initially motivated by several different observations and early simulations \citep{pravec2000,richardson_flyby}.
If we start with the idea that asteroids are aggregates of different sized components, and not singular monolithic bodies, it is possible to study them with some of the tools that have been used in the fields of Soil Mechanics and Granular Dynamics. In them, parameters such as porosity, cohesive and tensile strength, angles of friction and repose, particle size distributions, stress states, heterogeneity and yield criteria among others, determine how these granular systems will react when subjected to different, changing, external factors. These external factors are believed to have produced and shaped the asteroids that now exist around us and include solar photon momentum, gravitational tides, micro- and macro-impacts.
In this paper we will review what is known about the surface and interiors of granular asteroids, how different theoretical, experimental and simulation tools have been used to study them, how space missions and ground-based observations have shaped our understanding of their physical reality, and what we expect to learn from future missions. This article will also touch on some of the latest findings obtained by different groups. In particular we will discuss the rotational evolution of self-gravitating aggregates under the influence of the YORP effect and how their angles of friction, tensile strength, porosity, internal structure and density give rise to different disruption modes and the role they play in the formation of asteroids pairs, tumblers and binary systems.

\section{Granular Asteroids}

The first thing that we need to define is the name that we will use to call these self-gravitating aggregates, gravitational aggregates or ``rubble-pile'' asteroids.  In every case, when a researcher refers to this particular kind of asteroids, which are not monoliths, but an aggregate, they are talking about an asteroid which is made up a number of solid particles that are bound together, as a solid body, by gravitational and possibly cohesive/adhesive forces.  From the images and sample obtained by the Hayabusa mission to asteroid Itokawa, the disruption events of active asteroids P/2013 R3 and P2013 P5 and many other observations, we know that the particles that form asteroids have a size distribution that goes from $10^{-6}$ to tens of meters, and therefore from dust, boulders and everything in between.

This description is akin to the simple and clear definition for granular materials given by \cite{jaeger}: ``Granular materials are simple: they are large conglomerations of discrete macroscopic particles. If they are noncohesive, then the forces between them are only repulsive so that the shape of the material is determined by external boundaries and gravity.''  Comparing both concepts then, it is easy to see that they both refer to the same kind of matter, granular matter.  The only difference being that the authors of the latter did not think of including asteroids, or self-gravity that makes redundant the use of external boundaries, in the definition.  Taking these two ideas into account, in what follows, instead of using the common term ``rubble-pile''  asteroids, we will use the term {\it granular asteroids} as it seems much more appropriate, descriptive, general and accurate.

\section{A Brief Introduction to Granular Materials}

As stated above, granular materials are simple, in fact human kind has been dealing with them for thousands of years.  As stated by \cite{richard2005} ``Granular materials are ubiquitous in nature and are the second-most manipulated material in industry.''  One of the main characteristics of granular materials is the dissipative nature of their interactions.  In every collision there is always deformation and friction that contribute to the dissipation of energy.  This is the reason why a single marble dropped on the floor will bounce for a few seconds, but a bag full of them, when dropped, would have a dead stop.  Almost all of our food supply (fruits, grains and vegetables) are in a granular state; the motion of grains of sand, human and vehicular traffic can be understood as the motion of granular matter.  The formation of dunes in beaches and desserts on Earth as on other planets \citep{parteli2014}, as well as the structures found in planetary rings have been studied as granular systems.  They all share the same characteristics, they are formed by macroscopic particles that interact through long or short range forces and which collisions are highly dissipative.

\subsection{Friction and Cohesion}

The concepts of friction and cohesion in a granular aggregate have their origin in the inter-particle surface forces between individual grains, but they are not the same.  Both forces have an electromagnetic origin and appear as a result of the interaction between the atoms that form the surfaces of any two bodies \citep{hamaker1937, makkonen2012}.  In the context of a granular aggregate, friction must be understood as the opposition of a body to deform under shear stress, opposition that is the result of frictional forces and geometrical interlocking.  Stress analysis can prove that an angle of internal friction is defined as the ratio of shear to normal stress when a granular media deforms under pressure.  In the same context, cohesion (or cohesive strength) is defined as the shear stress at zero normal stress in the Mohr-Coulomb stress criteria.  In turn, tensile strength, which is much more directly related to cohesive forces, is defined as the normal stress at zero shear stress.

\subsection{Simulation Methods}

Most of the simulations methods that have been applied in the research of granular asteroids have not been developed in the Planetary sciences community, but in the field of Soil Mechanics and Granular Dynamics.  Among these methods we have:  Soft-Sphere Discrete Element Method (SSDEM) \citep{cundall1979, sanchez2011,sanchez2012,schwartz_SSDEM}, Hard-Sphere Discrete Element Method (HSDEM) \citep{alder1959,Richardson1994}, Contact Dynamics (CD) \citep{moreau1994}, the TC model \footnote{Though not directly stated, the TC model appears to have been named after the notation used to define the duration of a contact, {\it $t_c$}.} \citep{luding1998}, Finite Elements Method (FEM) and Lattice Element Method (LEM) \citep{affes-LEM3D}.  From them , SSDEM, HSDEM and FEM have been used very successfully in the simulation of asteroids.

{\bf SSDEM:}  Each particle is simulated as a separate entity that interacts with the others via a soft potential; therefore its name.  Particles are usually simulated as spheres for the simplicity of the collision detection algorithms, but others shapes are also possible.  This method is commonly referred to as Molecular Dynamics (MD) as it has been developed and used by different groups in different scientific communities.  Typically, the forces involved in simulations of granular media are short-range interactions, be these normal contact forces or surface-surface friction.  This makes possible the use of simple linked-cell algorithms to detect collisions \citep{allen}.  However, gravitational or electrostatic, long-range forces, can and have also being implemented, but they have required more complex algorithms for their calculation \citep{sanchez2012}.

{\bf HSDEM} Here too, each particle is simulated as an individual entity, but the particle-particle interactions are not carried out through forces but as a momentum exchange that responds to a pre-set coefficient of normal or tangential restitution.  The collisions can be only binary, as they are idealized as instantaneous, making this method ideal for highly kinetically active systems where the time that a particle needs to cover the mean free path is much longer than the duration of a collision.  However, it is not recomended for systems in a condensed phase where particles sustain lasting contacts as this is avoided by construction.  Here, the collisions have to be calculated hierarchically in time.  The TC model was developed in order to overcome some of the shortcomings of this method \citep{luding1998a}.

{\bf FEM} In this method, the size of the system is considered much larger than the size of any individual component and so it is studied as a continuous media.  Characteristics as density, porosity, stress fields, angle of friction, cohesive and tensile stress together with flow rules and yield criteria define how this media will react to different factors \citep{holsapple, hirabayashi2014}.

\section{Evolution of Granular Asteroids}

The evolution of granular asteroids has to do with their change in response to external factors.
These external factors produce events that form, deform or disrupt the structure of these asteroids in short or long periods of time.  The structure of a granular asteroid is defined, as in any other soil, by: size, shape, density, local gravity, cohesion, angle of internal friction, porosity and internal stress fields among other factors \citep{holsapple2001, holsapple, holsapple2010, sanchez2012, sanchez2014, hirabayashi2014}.

Among the external factors that affect the evolution of granular asteroids we have: gravitational tides, impacts, solar photon momentum.  In turn, their influence will be affected by the location of the asteroid in the Solar system as well as its spin properties.  This will be the topic of the coming sections, a brief review of the current understanding of how all these external effects have interplayed with the structure of asteroids to provoke their evolution.  In this paper, fission is defined as {\it any event in which a portion of the main aggregate separates from it, regardless of their mass ratio} \citep{scheeres_fission}.

\section{Past Work}

\subsection{Simulations}

The first works that treated asteroids as self-gravitating granular media were carried out by the groups lead by Prof. Dereck Richardson and Dr. Patrick Michel (e.g. \citep{Richardson1994, leinhardt, DCR_AIII, Richardson_tidal}).  The entirety of their computational research was carried out using PKDgrav, an N-body code that originally implemented only the HSDEM that, as explained above, is ill suited for aggregates in a condensed phase \citep{allen, mitarai2003}.

Among the many contributions carried out with the HSDEM implementation, there are some notable works.  The research carried out by \cite{richardson_flyby} studied the dynamics of granular asteroids subjected to tidal distortions and found links to: crater chain formation on the Moon from rubble pile fragments, formation of binary asteroids and doublet craters, Earth crossing asteroids (ECAs) with satellites, ECAs with irregular shapes, and size and orbital distribution of ECAs.  On the other hand, the research presented by \cite{richardson_impact} highlighted the results of collisions (normal and oblique) between granular asteroids.  Their results showed that: larger impact angles result in more elongated, faster-spinning remnants; larger impact speeds result in greater mass loss and increased mixing of the remnant; and initial impactor spin can increase or reduce the rotation period and elongation of the remnant.

Of course, we could not continue this review without mentioning the influential work carried out by \cite{walsh_binary} on the formation of binary asteroids at high spin rates; which was expanded in \cite{walsh2012}.  This research showed how the surface of granular asteroids could flow, form an equatorial ridge and eject material that will subsequently re-accrete to form a satellite that would orbit around the original body.  However, their high angles of friction ($\approx 40^{\circ}$) could be obtained only when the particles of the aggregates were crystalized.

Also among their various works, they started to look into the effect that two different cohesive strength models (rigid and elastic) could have in the dynamics of their simulated aggregates \citep{richardson2009}.  However, here they found that for the rigid case, no binary system was formed, whereas for the elastic case, some aggregates formed a satellite.  At the time, these two strength models were in development, but no further update was made in the HSDEM version of PKDgrav.

The introduction of the Soft-Sphere Discrete Element Method (SSDEM) for modelling of granular asteroids was made by \cite{sanchez-dps2009,sanchez2011, sanchez2012}.  Following this the PKDgrav program was also modified into an SSDEM code in \cite{schwartz_SSDEM} and since then others have also started to use this version of the code for their research; examples of this are \cite{yu2014} and \cite{ballouz2015}.  The former deals with the collisional disruption of rotating granular aggregates whereas the latter studies the possible effects on the surface of asteroid 99942 Apophis during the 2029 close approach.

\subsection{Theoretical Work}

In addition to the simulation efforts above explained, theoretical work was also carried out.  Among the main contributors we could cite the works of Keith Holsapple (U. of Washington), Ishan Sharma (IIT Kampur), Daniel Scheeres and Masatoshi Hirabayashi (U. of Colorado Boulder).  The main characteristic of their research was, and is, the use of theoretical tools commonly used in the field of soil Mechanics in order to understand the deformation processes and failure modes of granular asteroids.

Among the many outcomes that came from this approach it was found that the common tools used for soil mechanics were indeed applicable to granular asteroids.  Also, the sole knowledge of the shapes of Solar System bodies was not enough to claim that the bodies must be cohesive; they were almost all consistent with a granular, moderate porosity structure \citep{holsapple2001}.  \cite{sharma2009} continues the work on cohesionless granular asteroids and presents a unified method to characterize the equilibrium shapes, and passage into such states, of ellipsoidal asteroids with interiors modeled as cohesionless, rigid-perfectly-plastic materials following a Drucker-Prager yield criterion.

However, \cite{holsapple2007} found that the presence of tensile and cohesive strength for a large body ($>$10 km) makes no difference in the permissible spin because for such large sizes, the strength in a monolithic body is expected to be very small compared to self-gravity. Therefore, the observed spin limit for large bodies cannot be used to infer zero-strength (cohesive/tensile) granular bodies. On the other hand, the strength that allows the higher spins of the fast rotators (km size bodies and less) is only on the order of 10$-$100 kPa, which is a very small value compared to the strength of small terrestrial rocks. So these bodies do not need to be very strong, they could be essentially aggregates that have accumulated slight inter-particle bonding.  This of course, opens a question about the source of the strength of granular asteroids.

\cite{scheeres_fission} uses the energetics and dynamics of contact binary asteroids as they approach and pass the rotational fission limit to explain the spin-rate barrier, contact binaries, and the observed morphology of most NEO binaries.  This study is next complemented by \cite{jacobson-ApJ2011}.

\section{Present Work}

\subsection{The Source of the Strength of Granular Asteroids}

Though much progress had already been made by previous research, the asteroid-size vs. spin rate presented by \cite{pravec2000} still needed to be explained.  Theory had explained that, if not monolithic, asteroids had to have some cohesion to stay intact at the elevated spin rates that could be observed.  \cite{Icarus_scaling} put the different forces that could be present in an asteroid into context with micro-gravity field.  This led to the conclusion that, although cohesive van der Waals forces are very weak, in a small asteroid, they may turn out to be as important as gravity.  One prediction of this model was that small asteroids may be made of an assemblage of monolithic boulders as well as cohesive gravels of small enough size.  The problem with trying to prove this prediction is that the actual environment of a small asteroid cannot be reproduced in a laboratory.  Therefore, the only path to follow is simulations.  Here the problem is different, the large amount of particles that are needed to recreate the size distribution of a real small, granular asteroid is impossible to achieve with the available computational resources.  A  cohesive force can always be added to the contact law in simulations \citep{schwartz2013}, but it is necessary to understand its physical origin if we want to understand how asteroids evolve.  Only by doing this, simulations can be used to understand real physical systems.

\subsection{Simulating Cohesion}

\cite{sanchez2014} tackle the problem of cohesive strength in granular asteroids by focusing in the micro-mechanics that produces it instead of just adding a cohesive force.  They use a SSDEM code to simulate the granular equivalent to a liquid bridge, a granular bridge between two 1m spheres (boulders).  This bridge was formed by thousands of cm size particles (regolith).  The experiments consisted in pulling the two boulders apart until the bridge broke as a way to measure its tensile strength.  When the particles where cohesionless, the bridge broke as soon as the pulling force exceeded the self-gravitating forces.  On the other hand, when the particles where cohesive (van der Waals forces) the net pulling force needed to break the bridge was found to be inversely related to the average regolith size \footnote{The tensile strength is also related to other parameters such as the Hamaker constant of the material, packing fraction and directionality of the particle-particle contacts.}.  This is, the smaller the regolith, the greater the tensile strength of the bridge.  Additionally, it was also found that the bridge would break in a brittle fashion.

This finding was subsequently used to propose a model in which the smaller regolith and dust that would be present in granular asteroids \citep{science_Eros,thomas_eros_craters,science_nakamura2011, tsuchiyama2011, miyamoto_itokawa} could form a sort of ``van der Waals cement'' that could bind the bigger rocks and boulders in place.  This would allow km-size asteroids to rotate at the observed spin rates without disrupting.  Corroboration of this model would come from the analysis of the disruption of active asteroids (main belt comets) P/2013 R3 and P/2013 \citep{jewitt2013, jewitt2014}, and also from an analysis of the dynamics of asteroid 1950 DA \citep{rozitis2014, hirabayashi2014d}.  Interestingly enough, the research of \cite{hirabayashi2014d} also showed that 1950 DA may have a weak interior due to the shearing produced by its deformation.

\subsection{Disruption Patterns}

With these results at hand, it was observed that simulations of self-gravitating aggregates presented different disruption patterns at the critical spin rate \citep{walsh_binary, tanga2009, sanchez2012} depending on the numerical techniques and code, as well as other assumptions.  Particle-by-particle fission as well as the fission of coherent groups of particles were observed to occur in yet very similar simulations of non-spherical aggregates.  Different angles of friction and different particle packing (crystalline vs. random) could have been partially responsible for the difference in the outcome.  This showed that the friction angle was a factor in the disruption patterns of aggregates that needed to be systematically studied.  Additionally, cohesion was seen as important for the evolution of small asteroids and this was another parameter that needed to be probed.  With this in mind \cite{sanchez-acm2014} carried out a survey on ellipsoidal and spherical self-gravitating aggregates  with three different angles of friction (12$^\circ$, 25$^\circ$ and 35$^\circ$) and different values for cohesive strength (the ratio of cohesive to tensile strength is the tangent of the angle of friction in the Mohr-Coulomb yield criterion) that were spun to disruption rates and observed their behavior.  These simulations showed some interesting features: 1. at the lowest angle of friction, there is always great deformation and then the formation of a tail of particles at disruption spin rates, 2. the greater the angle of friction the smaller the allowed deformation of the aggregate, 3. for spherical aggregates (35$^\circ$), disruption was initially a particle-by-particle fission (as observed by \cite{walsh_binary,walsh2012}) that would slowly evolve to the fission of large coherent groups of particles, 4. at the highest cohesion, the aggregate was always divided in two.  This meant that the size of the fissioned piece was directly related to the cohesion of the aggregate.

Some of the simulations showed the formation of a system with a large primary and a small ``satellite.''  Though a rigorous analysis is still lacking to establish whether or not this was indeed a binary system, if this is so, it would be possible to propose a model that does not require the re-accumulation of fissioned  material from the primary to form the secondary.  This would be possible if the secondary detached from the primary as a coherent body, fully formed.  A result that has to be highlighted was that the shape of the primary is similar to asteroid 1999 KW4 alpha, though more stretched.  Of course, if simulations show great similarity to reality, this is a good result, but the question remains: why not a perfect fit?  The reason could be in the general assumption that theory and simulations have made that asteroids interiors are homogeneous.  Something that, if we look at the images taken by different space missions of the asteroid surfaces, is evidently very unlikely.  Additionally, and possible more importantly, in reality, the parameter space is vast, and little initial differences can lead to great differences in the outcome. All these geophysical processes are highly non linear so we can just hope to find qualitative fits, but not perfectly quantitative ones, and still be correct in our understanding.

\subsection{Heterogeneous Asteroids}

With the results outlined above, and if a better fit for asteroid 1999 KW4 alpha was pursued, it was clear that the strength of the aggregate needed to be increased, but at the same time the size of the detached piece had to be kept unchanged so that the size ratio of a binary system could be preserved.  The solution of this conundrum is simple, the aggregate required a stronger interior, a core.  This would supply the desired strength to avoid an excessive deformation, at the same time maintaining the weaker surface from which the secondary could be formed.  Also, the theoretical work carried out by \cite{hirabayashi2014b} and \cite{scheeres2015} showed that such a model could reproduce the shape profile of the above mentioned asteroid.  Another outcome of this research was that a homogeneous self-gravitating aggregate would fail at the centre first at the critical spin rate.  This was also observed by \cite{sanchez2012}.

\begin{figure}[htbp]
\begin{center}
\includegraphics[scale=0.55]{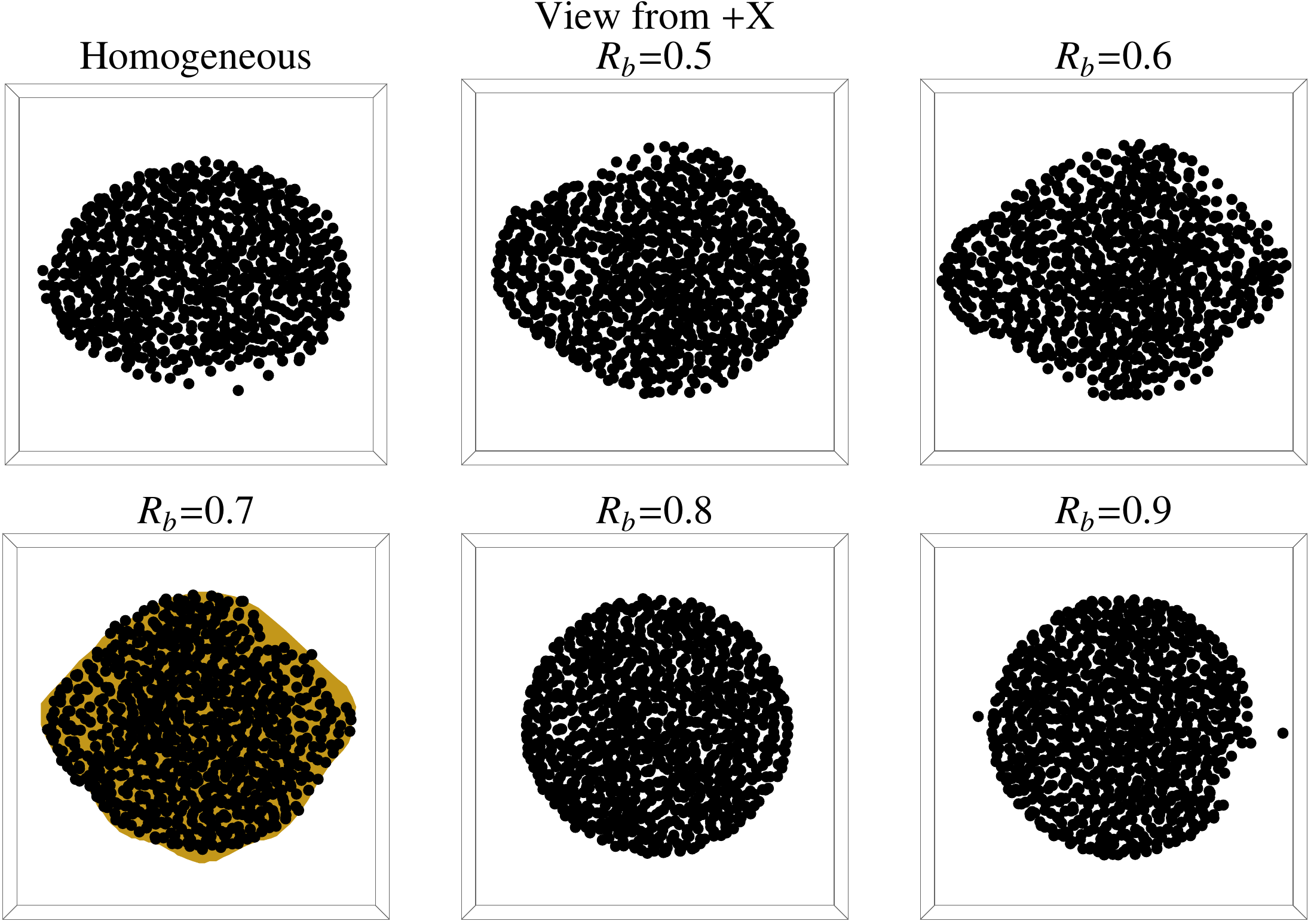}
\caption{ Cross-sectional front view of six aggregates with different size, strong cores.  $R_b$ is the normalized radius of the core with respect to the total size of the original aggregate.  The coloured shape that serves as a background for the image at the lower-left corner corresponds to asteroid 1999 KW4-alpha \citep{shapemodelsJPL}.}
\label{xview}
\end{center}
\end{figure}

\cite{hirabayashi2015} carried out theoretical work and simulations to show the deformation and disruption dynamics of such systems.  They arrived to the conclusion that for a granular asteroid to disrupt through surface shedding, it needed to have a strong core.  This could imply that the difference in disruption patterns observed in P/2013 R3 and P/2013 P5 is due to the differences in the internal structures.  Figure \ref{xview} shows the profile of the simulated aggregates that were used by \cite{hirabayashi2015}.  In them, $R_b$ is the radius of the core normalized to the radius of the aggregate.  These images show that the existence of a strong core inhibits the deformation caused by rotation.  Also, and more importantly, the shape of the aggregate in the lower left corner ($R_b=0.7$) is remarkably similar to 1999 KW4 alpha.  As a matter of fact, if we were to overlap an image of its shape model, the fit would be almost perfect.

\begin{figure}[htbp]
\begin{center}
\includegraphics{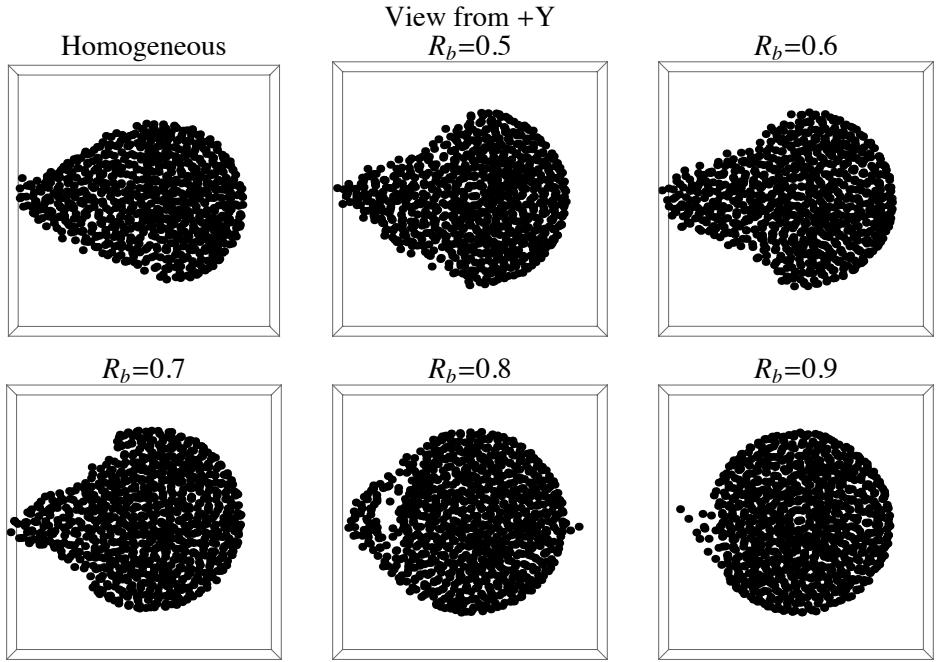}
\caption{ Cross-sectional side view of six aggregates with different size, strong cores.  $R_b$ is the normalized radius of the core with respect to the total size of the original aggregate}
\label{yview}
\end{center}
\end{figure}

Of course, this is the view from one particular perspective.  A look from the side would reveal that the match is not as good.  Fig.~\ref{yview} shows a side view of these aggregates and the aggregate with $R_b=0.7$ shows a very jagged profile.  If $R_b=0.8$ however, the shape has a softer profile; closer, but not matching the one of 1999 KW4.  These discrepancies and similarities only point to the simplicity of the model.  Obviously asteroids do not need to start from an almost perfect spherical shape and, if they have a core, it does not have to be concentric and spherical.  However, this result is very encouraging and means that we are in the right path.  Can we say something about 1999 KW4?  Maybe just one thing, it is probable that it has a core, possible with a size of $\approx70\%$ of the size of the asteroid.  This also agrees with the findings of \cite{walsh2012} in which they found that aggregates with solid cores close to this size would form binary systems more frequently.

There is also another direction that we could take and that is of aggregates with a weak interior.  This is still work in progress and will be reported elsewhere.

\section{Conclusions}

This paper is in part a brief review of the findings of different research groups on the evolution of granular asteroids.  These results have been obtained through simulations, theory and even experiments when possible.  There are some things we are beginning to understand, but if Granular Dynamics was a new field about 25 years ago for Physicists, low-gravity is a completely new realm.  Many questions are still open, among them: the reason behind particulate size segregation, the role of particle shape, other forms of heterogeneities, the role of fractures, super-fast rotators, etc.  There is still much work to be done and not many doing it.

Better and faster simulations are still needed and the work should not be entrusted only to a handful of people.  Meaning that students should begin to implement and work with their own codes, with new and more reliable complex methods that help us to tackle the limitations of our current tools.  Better simulations come from a greater understanding of the theoretical framework and so theoretical efforts are not only wanted but required.

Experimentation is still the only way to prove any and all the theories that have been put forward; however, there is still no easy way to mimic self-gravity in the laboratory or to witness processes that take millions of years.  In spite of this, sub-orbital and parabolic flights are being used to better understand the microgravity environment \citep{murdoch2013a}.  Additionally, the OSIRIS-REx (NASA) and Hayabusa2 (JAXA) space missions will bring valuable, new data that will help us to verify or refine our current understanding.  We eagerly await this to happen, this will get us one step closer to understand what once was called {\it the vermin of the sky}.

\section{Acknowledgements}

I gratefully acknowledge the support from NASA grant  NNX14AL16G from the Near-Earth Object Observation Program and from grants from NASA's SSERVI Institute.  

\bibliographystyle{my-iau}
\bibliography{/Users/paul/Documents/UCB/Meeting/psbib}

\end{document}